\documentclass[%
 reprint,
 amsmath,amssymb,
 aps,
]{revtex4-2}

\usepackage{xcolor}
\usepackage{graphicx}
\usepackage{dcolumn}
\usepackage{bm}
\usepackage{hyperref}
\usepackage{dutchcal}

\usepackage{multirow, hhline} 

\definecolor{darkgreen}{RGB}{0, 130, 33}

\newcommand{\proof}[1]{{\color{black}#1}}

\begin{document}

\title{Sky marginalization in black hole spectroscopy and tests of the area theorem}

\author{Alex Correia}
 \affiliation{University of Massachusetts Dartmouth, 285 Old Westport Rd, North Dartmouth, Massachusetts 02747, USA}
\author{Collin D. Capano}%
 \email{cdcapano@syr.edu}
\affiliation{Department of Physics, Syracuse University, Syracuse, New York 13244, USA}
\affiliation{University of Massachusetts Dartmouth, 285 Old Westport Rd, North Dartmouth, Massachusetts 02747, USA}

\begin{abstract}
Direct observation of gravitational waves from binary black hole (BBH) mergers has made it possible to test the laws of black hole thermodynamics using real astrophysical sources. These tests rely on accurate and unbiased parameter estimates from the pre and postmerger portions of a signal. Due to numerical complications, previous analyses have fixed the sky location and coalescence time when independently estimating the parameters of the pre and postmerger signal. Here we overcome the numerical complications and present a novel method of marginalizing over sky location and coalescence time. Doing so, we find that it is not possible to model only the pre or postmerger portions of the signal while marginalizing over timing uncertainty. We surmount this problem by simultaneously yet independently modeling the pre and postmerger signal, with only the sky location and coalescence time being shared between the models. This allows us to marginalize over all parameters. We use our method to measure the change in area $\Delta A_{\rm measured} = A_f - A_i$ between the final and initial black holes in the BBH merger GW150914. To measure the final black hole's area $A_f$ we do an analysis using quasinormal modes (QNMs) to model the postmerger signal, and another analysis using the postmerger portion of an inspiral-merger-ringdown (IMR) template. We find excellent agreement with expectations from general relativity. The Hawking area theorem (which states that $A_f \geq A_i$) is confirmed to $95.4\%$ and $99.5\%$ confidence using the QNM and IMR postmerger models, respectively. Both models yield $\Delta A_{\rm measured} / \Delta A_{\rm expected} \sim 1$, where $\Delta A_{\rm expected}$ is the expected change in area derived from fits to numerical relativity simulations.
\end{abstract}

\maketitle

\section{Introduction}

The direct detection of gravitational waves (GWs) throughout the past decade has been one of the most significant advancements in observational relativity. These observations not only confirmed the existence of GWs, but also opened the door to direct tests of general relativity (GR) in the strong-field regime. One such prediction that can be verified is Hawking’s area theorem, which states that the remnant from a binary black hole (BBH) merger must have an event horizon surface area greater than the sum of the progenitor horizon areas \cite{area-theorem}.

Tests of the area theorem using GWs have been carried out previously~\cite{kastha-area-thrm, cabero-area-thrm, isi-area-thrm}. Generally, they involve measuring the initial two black holes’ mass and spin (from which the total initial area $A_i$ is derived) from the premerger part of the signal, during which the two black holes inspiral into each other. The area of the final black hole is independently measured using the GW that is emitted during the post merger, or ``ringdown'' phase. In both cases Bayesian inference is used to produce a ``posterior'' probability density on the black holes' parameters. Ideally, all parameters that describe the BBH should be allowed to vary in the analysis, in order to fully account for all statistical uncertainties. However, previous tests of the area theorem have fixed the sky position and coalescence time t c of the events to nominal values when doing their analysis~\cite{kastha-area-thrm, isi-area-thrm}. Fixing the values in this way may lead to biases in the resultant parameter estimates, obfuscating the true nature of the system \cite{cabero-area-thrm, Wang:2023xsy}. At the very least, it may cause an underestimate of the statistical uncertainty of measured parameters, yielding constraints on deviations from GR that are misleadingly strong.

The sky location and $t_c$ have been fixed in earlier studies due to technical hurdles in calculating the likelihood function. In order to independently analyze the pre and postmerger portion of the signal it is necessary to excise the post and premerger data, respectively, from the analysis. Excising the premerger data is also necessary when just analyzing the postmerger signal for tests of the no-hair theorem using quasinormal modes (known as black hole spectroscopy). In either case, the excision (or ``gate'') results in a modified likelihood function that cannot be solved using conventional, frequency-domain means. Existing pipelines for doing such analyses in the time domain, such as \texttt{pyring} \cite{pyring-methods, pyring-repo} and \texttt{ringdown} \cite{isi-ringdown}, instead calculate the likelihood by numerically inverting the noise covariance matrix for the data. This calculation can be numerically costly~\cite{Isi:2021iql,Wang:2023xsy}. If the sky location and $t_c$ are not fixed, the gate will vary in time, and the full likelihood will need to be recalculated for every unique gate time, significantly increasing computational costs. However, if the sky location and $t_c$ are fixed, the gate is static, and the most computationally demanding part of the calculation need only be done once.

Finch and Moore \cite{finch-moore} devised a method to overcome this issue and vary the sky location and $t_c$. In their method the inspiral and merger are modeled with a wavelet series stitched to the beginning of the ringdown. This allows them to use the traditional frequency-domain likelihood, and vary the start time of the ringdown. However, this does not allow for an area theorem test, since the parameters of the initial black holes cannot be estimated from the wavelet signal model. Furthermore, the traditional frequency-domain likelihood causes the start of the ringdown to be coupled to information from just before the merger due to convolution with the whitening filter. This means that recovered pre and postmerger parameters will not be independent measurements.

This paper presents a novel method of calculating the likelihood function using the parameter estimation code \texttt{PyCBC Inference} \cite{pycbc}. This code already utilizes ``gating and in-painting" \cite{gating-inpainting} to excise data from the analysis, which allows for cheaper likelihood evaluation under certain conditions as compared to the methods used by \texttt{pyring} and \texttt{ringdown}~\cite{Isi:2021iql,Wang:2023xsy}. However, an additional normalization factor has traditionally been omitted from this calculation, as this also required computationally expensive numerical methods to evaluate. This paper presents a method of linear interpolation that calculates this normalization factor with good approximation. Together, these methods allow for fast likelihood calculations regardless of gate position, allowing for marginalization over $t_c$ and sky location and a full accounting of parameter uncertainties.

This paper is structured as follows. Section \ref{sec:det_approx} describes the modifications to the likelihood calculation in \texttt{PyCBC} in detail. Full waveform analyses can be conducted with these modifications to marginalize over sky location and $t_c$. However, the method cannot be used in partial waveform analyses; Section \ref{sec:challenges} documents the issues that arise in these models. Using this method, a test of the Hawking area theorem is conducted using data from GW150914 \cite{GW150914-observation}. Section \ref{sec:methods} describes the configuration of the analyses used in this test, and Sec.~\ref{sec:area_thrm} describes and analyzes the results. Section \ref{sec:conclusions} summarizes the findings and the implications of the marginalization method for future GW analyses.

\section{Determinant approximation} \label{sec:det_approx}

\texttt{PyCBC Inference} \cite{pycbc} utilizes Bayesian inference to conduct parameter estimation on GW events. Bayes' theorem is used to extract information about the parameter space $\boldsymbol{\vartheta}$ from a dataset $\boldsymbol{s}$ \cite{bayes-thrm}. Assuming the dataset is composed of a signal $h$ and noise $n$ such that $\boldsymbol{s} = \boldsymbol{n} + \boldsymbol{h}$, the probability of observing a specific $\boldsymbol{\vartheta}$ is given by
\begin{equation}
    p(\boldsymbol{\vartheta} | \boldsymbol{s}, h) = \frac{p(\boldsymbol{s} | \boldsymbol{\vartheta}, h)p(\boldsymbol{\vartheta} | h)}{p(\boldsymbol{s} | h)}.
    \label{eqn:bayes_thrm}
\end{equation}
The term $p(\boldsymbol{s} | \boldsymbol{\vartheta}, h)$ is the likelihood, which describes the probability of observing a signal $\boldsymbol{s}$ assuming the event has a parameter space $\boldsymbol{\vartheta}$. The term $p(\boldsymbol{\vartheta} | h)$ is the prior, a distribution that describes the \textit{a priori} probability of observing $\boldsymbol{\vartheta}$ given a signal model $h$. The denominator $p(\boldsymbol{s} | h)$ is the evidence, a normalization factor used to compare analyses using different models for $h$. The resultant probability distribution on the left-hand side of the equation is known as the posterior.

The prior is chosen at the discretion of the analyst based on assumed plausible values for $\boldsymbol{\vartheta}$, whereas the likelihood is calculated directly from the data. For a system of $K$ detectors each taking $N$ time series samples of an event with a stochastic Gaussian noise background, the likelihood can be written as \cite{likelihood}
\begin{equation}
    p(\boldsymbol{s}_{net} | n) = \frac{\exp{[-\frac{1}{2} \sum_{d=1}^K \boldsymbol{s}^T_d \boldsymbol{\Sigma}^{-1}_d \boldsymbol{s}_d]}}{[(2\pi)^{NK} \prod_{d=1}^K \det \boldsymbol{\Sigma}_d]^{1/2}},
    \label{eqn:det_likelihood_app}
\end{equation}
where $\boldsymbol{\Sigma}_{d}$ is the covariance matrix associated with $n$ in detector $d$.

To evaluate the likelihood, further assumptions must be made to calculate $\boldsymbol{\Sigma}_{d}^{-1}$. The result of these assumptions is
\begin{equation}
    \label{eqn:noise_likelihood}
    p(\boldsymbol{s}_{net} | n) \propto \exp \bigg[-\frac{1}{2} \sum^K_{d=1} \langle \boldsymbol{s}_d, \boldsymbol{s}_d \rangle \bigg],
\end{equation}
where the inner product is defined by Eq.~\eqref{eqn:inner_product}. (The full derivation of this likelihood function is given in Appendix \ref{sec:likelihood_der}). This gives the likelihood function for a signal that is assumed to be entirely noise. To get the likelihood function with respect to the signal model $h$ evaluated for a parameter space $\boldsymbol{\vartheta}$, Eq.~\eqref{eqn:noise_likelihood} can be rewritten by substituting $\boldsymbol{s}_d \rightarrow \boldsymbol{s}_d - \boldsymbol{h}_d (\boldsymbol{\vartheta})$ on the right-hand side, yielding
\begin{align}
    \log p(\boldsymbol{s}_{net} | \boldsymbol{\vartheta}, h) = &-\frac{1}{2} \sum^K_{d=1} \langle \boldsymbol{s}_d - \boldsymbol{h}_d (\boldsymbol{\vartheta}), \boldsymbol{s}_d - \boldsymbol{h}_d (\boldsymbol{\vartheta}) \rangle \nonumber \\ 
    &- \frac{1}{2}\bigg[ NK \log 2\pi + \sum^K_{d=1} \log \det \boldsymbol{\Sigma}_{d} \bigg].
    \label{eqn:loglikelihood}
\end{align}

(Here and throughout, we use $\log$ to refer to the natural logarithm.) For a GW analysis that examines the entirety of $\boldsymbol{s}$, this expression is easily evaluated using the approximated eigenvalues of $\boldsymbol{\Sigma}_d$. However, many analyses only examine a portion of the time series (generally either the pre or postmerger signal). Therefore, a region of $\boldsymbol{s}$ is omitted, or ``gated'', to reduce numerical biases due to the merger. Doing this also requires excising rows and columns from $\boldsymbol{\Sigma}_d$, thereby breaking the Toeplitz form of the matrix and the corresponding approximations. The inverse of the truncated covariance matrix $\boldsymbol{\Sigma}_{d,tr}$ is evaluated in \texttt{PyCBC} using ``gating and in-painting" (explained in detail in \cite{gating-inpainting} and Appendix \ref{sec:gate_paint}), allowing for easy calculation of the first term of Eq.~\eqref{eqn:loglikelihood}.

The second term, however, requires the calculation of $\log \det \boldsymbol{\Sigma}_{d,tr}$, which is a notoriously expensive numerical problem. The most widely used matrix decomposition methods are $\sim\!O(N^3)$ \cite{det-calc}, which become impractical for GW time series such as GW150914, where $N\!\sim\!10^4$. This problem is amplified when varying sky position and $t_c$ in a gated analysis. The start and end times of the gate must be converted between the geocentric and individual detector frames. The time conversions directly depend on the time of merger $t_c$ and the sky location of the event relative to the detectors. Therefore, when varying sky location and $t_c$, the gate times in the detector frames will also vary. Subsequently, the elements of $\boldsymbol{\Sigma}_{d,tr}$ will vary with each unique sky location and $t_c$ value. Analyses that marginalize over these parameters require recalculating $\log \det \boldsymbol{\Sigma}_{d,tr}$ on multiple steps of the sampler, which would be impractical using numerical decomposition methods. This is not an issue if the sky location and $t_c$ are fixed, since in that case $\det \boldsymbol{\Sigma}_{d,tr}$ will not change throughout the analysis (and in fact can be ignored, as it amounts to a constant normalization term).

\begin{figure}
    \centering
    \includegraphics[width=0.45\textwidth]{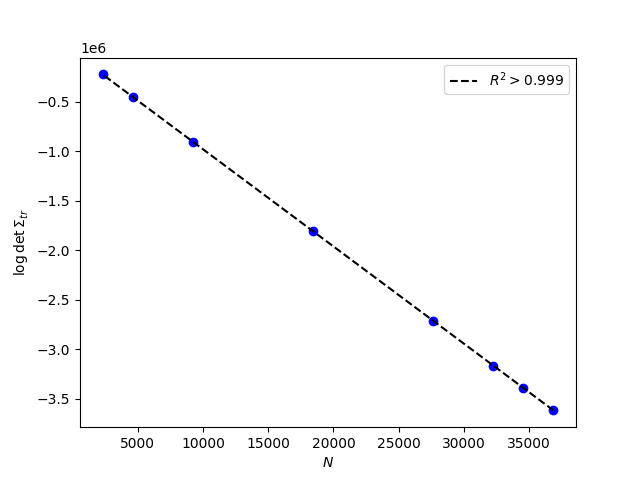}
    \caption{Plot of $\log \det \boldsymbol{\Sigma}_{d,tr}$ versus PSD length for GW150914 in the Hanford detector. Blue points represent determinants calculated exactly using \texttt{SciPy} functions \cite{scipy}, while the black dashed line is a least squares linear fit. The correlation coefficient of $0.999 < R^2 < 1$ indicates a highly significant linear correlation between $\log \det \boldsymbol{\Sigma}_{d,tr}$ and the size of $\boldsymbol{\Sigma}_{d,tr}$.}
    \label{fig:det_size_H1}
\end{figure}

In order to marginalize over sky location and $t_c$ it is necessary to have a fast method for calculating $\det \boldsymbol{\Sigma}_{d,tr}$. We find that $\log \det \boldsymbol{\Sigma}_{d,tr}$ is strongly linearly correlated to the length of the power spectral density (PSD) of the data, which is equivalent the number of rows and columns in $\boldsymbol{\Sigma}_{d,tr}$. Figure~\ref{fig:det_size_H1} shows this relationship using $\log \det \boldsymbol{\Sigma}_{d,tr}$ values calculated for various gate sizes applied to GW150914 data. Using a least squares fit to the points, the correlation coefficient was calculated as $0.999 < R^2 < 1$, which for a sample of eight points implies a highly significant correlation \cite{err-analysis}.

We also find that the position of the gate in the time series has no significant effect on the value of $\log \det \boldsymbol{\Sigma}_{d,tr}$. The maximum range over which the determinant values varied was $\sim\!10^{-6}$. However, since these values were generally of order $\sim\!10^6$, this represents a maximum fractional change of $\sim\!10^{-12}$ over the entire time domain. While gate position does have a minor effect on determinant value, the relative effect is so small that the value is well-approximated as a constant for a static gate length.

Using these two facts, we calculate the normalization term in Eq.~\eqref{eqn:loglikelihood} using a linear interpolation based solely on the size of $\boldsymbol{\Sigma}_{d,tr}$. Specifically, a least squares linear fit is generated using the determinant values for the full matrix and three differently-sized truncated matrices. The full matrix determinant is calculated using its approximated eigenvalues, while the truncated determinants are calculated numerically using \texttt{SciPy} \cite{scipy}. The determinant can then be calculated on each step through linear interpolation based on the size of the truncated covariance matrix.

\section{Need for Joint Analyses} \label{sec:challenges}

\begin{figure*}
    \centering
    \includegraphics[width=0.48\textwidth]{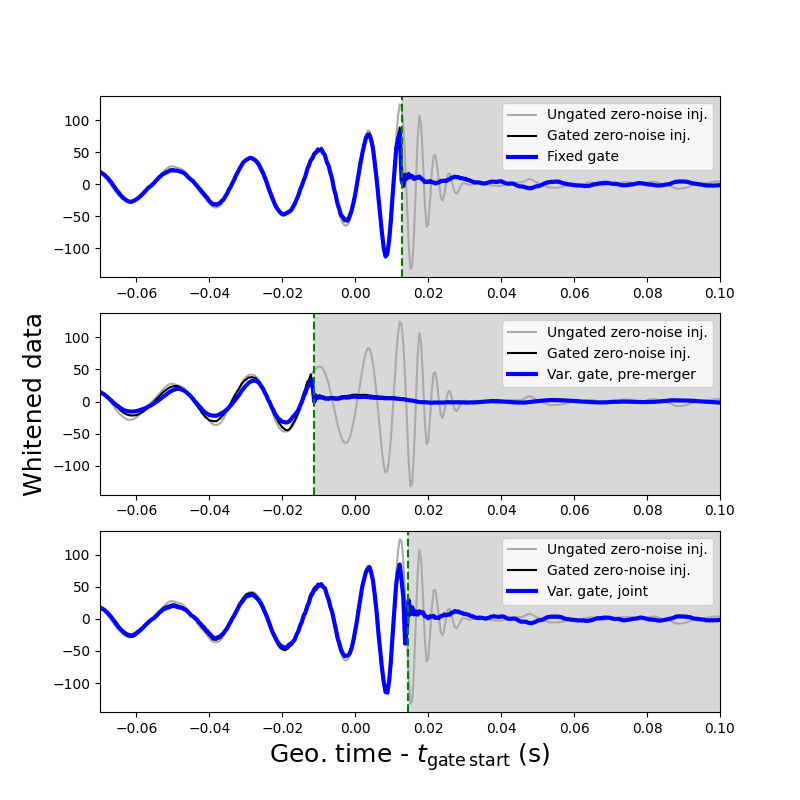}
    \includegraphics[width=0.48\textwidth]{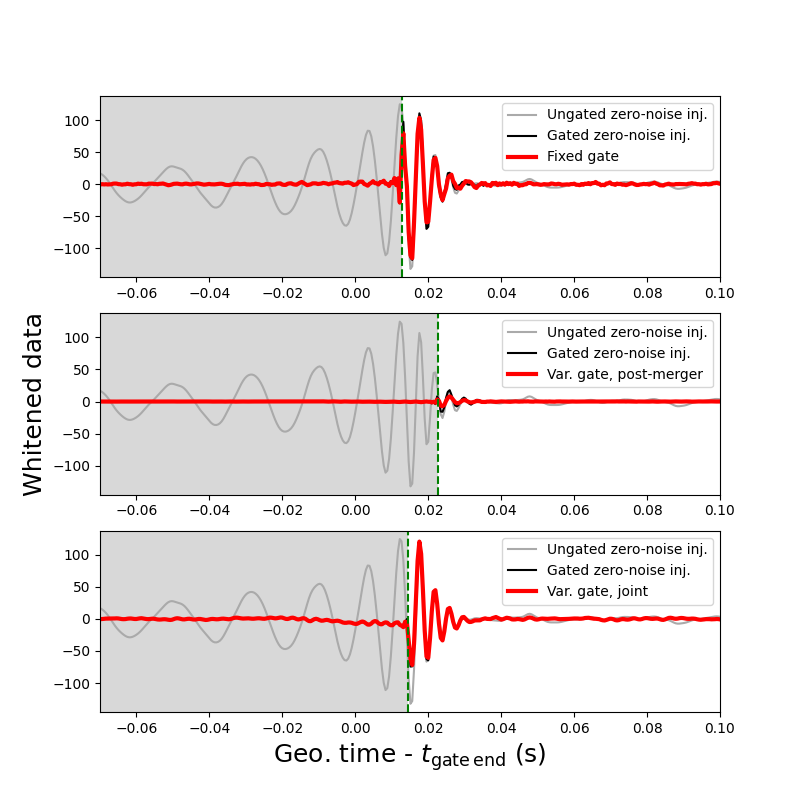}
    \caption{Comparison of waveform templates fitting to a simulated signal similar to GW150914 in the Hanford detector. The simulated signal (``injection'') is generated using the maximum likelihood values from a 4-OGC analysis of GW150914 \cite{4ogc} and added to zero noise. The ungated zero-noise injection is plotted with a gray line on each set of axes. The plots on the left (right) contain maximum likelihood premerger (postmerger) waveform templates plotted with blue (red) lines. In each analysis, a 1-second gate is applied to the data, represented by a shaded gray region and a dashed green line representing the start (end) of the gate. The data with this gate applied is shown with a black line. The horizontal axes represent the time in seconds relative to the geocentric gate start (end) time. The top plots show the waveform templates for analyses where the sky location was fixed to the same values as the injection. The center plots show the templates for analyses where the sky location was allowed to vary and the pre and postmerger signals were modeled separately. Notice that the premerger (postmerger) gate is significantly earlier (later) than the injected values shown in the corresponding fixed sky location templates. The bottom plots show the templates from an analysis with variable sky location where the pre and postmerger signals were modeled simultaneously. Besides sky location and $t_c$, the parameter distributions were independent of each other. The resultant gate positions are similar to those observed in the fixed sky location analyses.}
    \label{fig:wf_model_comp}
\end{figure*}

The strategy outlined in Sec.~\ref{sec:det_approx} fixes the technical hurdle of calculating the likelihood when the gate length and position are varied. Even with that, however, we find that it is not possible to only model a portion of a signal --- whether it be the postmerger or the premerger --- if $t_c$ or sky location are uncertain. This is because a larger likelihood can generally be obtained if the gate is shifted such that it removes as much of a signal as possible.

From Eq.~\eqref{eqn:loglikelihood} it is evident that the likelihood is maximized as $\boldsymbol{s}_d - \boldsymbol{h}_d (\boldsymbol{\vartheta}) \rightarrow 0$. When analyzing the entire time series this can only occur if the template $\boldsymbol{h}_d(\boldsymbol{\vartheta})$ is similar to the signal that exists in the data (as desired). However, if a variable gate is involved, then a large likelihood can also be obtained if the gate is shifted such that it excises most (or all) of the signal in the data. The template then only needs to match the remaining noise. Since the noise manifold is larger than the signal manifold, a larger volume of prior space will generally be able to match the noise than the volume that matches the signal. For example, when using a quasinormal mode (QNM) model, if the gate is shifted such that it excises the entire signal then one only needs to reduce the amplitude of the QNM template below noise level to get a relatively large likelihood. The same likelihood would then be obtained regardless of what the other parameters describing the template are set to. The end result is that the posterior probability will favor excising the entire signal even if the template that is being used can match the signal.

Specifically, in \emph{pre}-merger-only analyses, the gate will be shifted to as \emph{early} a time that is possible while in \emph{post}-merger-only analyses, the gate will be shifted to as \emph{late} as possible. These phenomena are illustrated in the middle row of Fig.~\ref{fig:wf_model_comp}. Shown are the results from separate premerger and postmerger analyses of a simulated signal in which the gate time is varied due to time and sky-location marginalization. The maximum likelihood template in the premerger-only analysis has a $t_c$ roughly $25 \text{ms}$ earlier than what was injected. This significantly shortens the premerger signal, leading to inaccurate measurements in parameters such as total spin. Meanwhile, the maximum likelihood template for the postmerger-only analysis had a ringdown start time $\sim\!10 M_{\odot}$ later than the injected value, well into the regime where the signal is expected to be noise dominated. This leads to a ``prior-in prior-out" posterior --- a roughly uniform distribution across parameter space --- as the signal model can fit any arbitrary model to late-time noise.

Fundamentally, this problem is due to the gated signal model not being the appropriate model for the observed data. Excising data from the analysis is mathematically equivalent to assuming that the excised data is Gaussian noise and marginalizing over all possible realizations of it. A property of multivariate Gaussian distributions is that marginalizing over a subset of dimensions yields another multivariate Gaussian distribution with the marginalized dimensions excised from the covariance matrix. This is exactly the same form as Eq.~\eqref{eqn:noise_likelihood}; in our case, each time sample is a dimension in the multivariate Gaussian. The problem is the excised times are \emph{not} Gaussian noise. They contain a signal, albeit a portion of the signal that we want to ignore. Consequently, a gated signal model is not representative of the data.

This issue could be mitigated by modifying the prior to exclude the portion of parameter space that matches the noise while keeping the portion that matches the signal. With a QNM template this would mean setting a lower bound on the amplitude such that it is above noise level. However, modifying the prior in this manner is challenging to do in an unbiased way, as it would require a priori knowledge of both the noise and signal properties. It also does not solve the fundamental problem that the gated signal model is not representative of the data.

We resolve this issue by simultaneously yet independently modeling both the pre and postmerger signals. Each domain is treated as a separate gated analysis. The signal parameters used in each domain are independent of each other, except for a set of common parameters $\boldsymbol{\vartheta}_{\rm com}$. In our case the common parameters are the right ascension $\alpha$, declination $\delta$, and geocentric coalescence time $t_c$. These parameters determine the coalescence time $t^d_c$ in each detector $d$, which the gate time depends on. The likelihood is calculated in each domain, then combined to get the overall likelihood.

Explicitly, our algorithm to calculate the likelihood for a given set of parameter values $\boldsymbol{\vartheta} = \{\boldsymbol{\vartheta}_{\rm insp}, \boldsymbol{\vartheta}_{\rm rd}, \boldsymbol{\vartheta}_{\rm com} \}$ is as follows:
\begin{enumerate}
    \item Generate the premerger (``inspiral'') template $h^{\rm insp}$ using parameters $\boldsymbol{\vartheta}_{\rm insp}$.
    \item Project the template into each detector's frame using $\boldsymbol{\vartheta}_{\rm com} = \{\alpha, \delta, t_c\}$. 
    \item Excise times after $t^{\rm det}_c$ by gating and in-painting the residual $s_d - h_d^{\rm insp}(\boldsymbol{\vartheta}_{\rm insp}, \boldsymbol{\vartheta}_{\rm com})$ using a $1\,$s gate that spans $[t^d_c, t^d_c+1)$.
    \item Calculate the premerger likelihood $p(\boldsymbol{s}_{net} | \boldsymbol{\vartheta}_{\rm insp}, \boldsymbol{\vartheta}_{\rm com}, h^{\rm insp})$ via Eq.~\eqref{eqn:loglikelihood}.
    \item Repeat steps 1--4 for the postmerger (``ringdown'') template $h^{\rm rd}$ to get the postmerger likelihood $p(\boldsymbol{s}_{net} | \boldsymbol{\vartheta}_{\rm rd}, \boldsymbol{\vartheta}_{\rm com}, h^{\rm rd})$. However, for the ringdown the gate spans $[t^d_c-1, t^d_c)$; i.e., it ends at $t^d_c$ whereas it starts at $t^d_c$ for the premerger. The data used in each domain is thereby (nearly) mutually exclusive (see Appendix \ref{sec:gate_paint} for more details).
    \item The total likelihood is then
    \begin{align}
        p(\boldsymbol{s}_{net} | \boldsymbol{\vartheta}, h) &= \nonumber \\
         p(\boldsymbol{s}_{net} &| \boldsymbol{\vartheta}_{\rm insp}, \boldsymbol{\vartheta}_{\rm com}, h^{\rm insp}) p(\boldsymbol{s}_{net} | \boldsymbol{\vartheta}_{\rm rd}, \boldsymbol{\vartheta}_{\rm com}, h^{\rm rd})
    \end{align}
\end{enumerate}

We fix the issue of the gate trying to excise the signal by using this hierarchical likelihood. The two domains offset each other; in order for the postmerger gate to shift to later times the premerger template must match more of the signal, and vice versa. This also addresses the fundamental issue with the gated signal model highlighted above; our global model for the entire dataset now contains a non-Gaussian element (the other domain's signal model) in each domain's excised region. Note also that no coupling occurs across the domain boundaries due to the whitening filter.

The bottom plots in Fig.~\ref{fig:wf_model_comp} show the maximum likelihood waveform templates from an analysis with this configuration. Besides sky location and $t_c$, both models generated independent parameter measurements. The resultant waveform templates closely match those obtained by fixing sky location and $t_c$ to the injected values (top row of Fig.~~\ref{fig:wf_model_comp}); the erroneous gate motion observed in the pre and postmerger-only analyses is no longer present.

\section{Methods} \label{sec:methods}

We performed eight analyses in total. Each analysis was differentiated by the waveform used, postmerger approximant, and whether or not sky location and $t_c$ were marginalized over. All other aspects of the analyses were kept constant. Both the pre and postmerger signals in all models utilized the PyCBC model \texttt{GatedGaussianMargPol}, which inherits the normalization protocols described in Sec.~\ref{sec:det_approx}. The \texttt{dynesty} sampler \cite{dynesty} was used to generate posterior distributions. The samplers in these analyses used 4000 live points to ensure the convergence of each model (see Appendix \ref{sec:addtl_results}).  

Half of the analyses used the original data of GW150914 with a sample rate of 2048 Hz obtained from the Gravitational Wave Open Science Center \cite{gwosc}. To validate these results, we repeat each run on a simulated signal in zero noise. The simulated signal was generated using the \texttt{IMRPhenomXPHM} waveform approximant~\cite{imrphenomxphm} with the maximum likelihood parameters for GW150914 from Ref.~\cite{4ogc}.

All template models utilized \texttt{IMRPhenomXPHM} (abbreviated here on as IMR) to model the premerger signal of the waveform. Half of the models used this IMR approximant to model the postmerger signal. The other half utilized a QNM approximant to model the postmerger section of the waveform. Table \ref{tab:inspiral_var_params} lists the sampled parameters and priors used in the premerger models, and Table \ref{tab:ringdown_var_params} lists the same for the postmerger models. The QNM approximant was configured such that the ringdown was composed of a dominant $(2,2,0)$ mode and a subdominant $(2,2,1)$ mode as proposed by \cite{isi-221-initial}. All models apply the ringdown model starting at merger time $t_c$. The priors of the QNM postmerger models were restricted such that the $(2,2,0)$ mode contribution to the postmerger signal-to-noise ratio (SNR) was at least 2. This condition was imposed to ensure that the dominant mode was present in the model, preventing possible ``label switching" that may occur due to the $(2,2,1)$ mode erroneously matching to the $(2,2,0)$ mode in the signal.

Only half of the models allowed for sky location and $t_c$ to vary. The other half fixed these parameters to nominal values for GW150914 to replicate previous works. Specifically, the fixed parameter analyses set $\alpha = 1.95$, $\delta = -1.27$, and $t_c = 1126259462.408$, in accordance with the maximum likelihood values in \cite{isi-221-initial}.

\begin{table*}
    \centering
    {\renewcommand{\arraystretch}{1.3}
    \begin{tabular}{||c|c||c|c||}
        \hhline{|t:==:tt:==:t|}
         Parameter & Description & Prior distribution & Prior range \\
        \hhline{|:==::==:|}
         $t_c$ & Coalescence time & Uniform & $1126259462.43 + [-0.05, 0.05]$ s \\
         \hhline{||--||--||}
         $\alpha$ & Right ascension & Uniform & $[0, 2\pi]$ \\
         \hhline{||--||--||}
         $\delta$ & Declination & Sine angle & $[-\pi/2, \pi/2]$\\
         \hhline{||--||--||}
         $\iota$ & Inclination & Sine angle & $[0, \pi]$ \\
         \hhline{||--||--||}
         $M_{chirp}$ & Source frame chirp mass & $M_1$, $M_2$ & $[23, 42] M_\odot$ \\
         \hhline{||--||--||}
         $q$ & Mass ratio $M_1/M_2$ & $M_1$, $M_2$ & $[1, 4]$ \\
         \hhline{||--||--||}
         $\chi_{a, (1/2)}$ & Spin magnitude & Uniform & $[0, 0.99]$ \\ 
         \hhline{||--||--||}
         $\chi_{\theta, (1/2)}$ & Spin polar angle & Solid angle & $[0, 2\pi]$ \\ 
         \hhline{||--||--||}
         $\chi_{\phi, (1/2)}$ & Spin azimuthal angle & Solid angle & $[0, \pi]$ \\ 
         \hhline{||--||--||}
         $\phi_c$ & Reference phase & Uniform & $[0, 2\pi]$ \\ 
         \hhline{||--||--||}
         $V_C$ & Comoving volume & Uniform & $[5000, 92918664351]$ Mpc$^3$ \\
         \hhline{|b:==:bb:==:b|}
    \end{tabular}}
    \caption{Varied parameters in IMR models and their associated prior distributions. The subscript $(1/2)$ indicates that the same prior was used for the primary and secondary masses. The third column indicates the sampling method for each prior. Parameters listed in this column indicate uniform sampling over those parameters rather than what is listed in column 1. (For example, $M_{chirp}$ is sampled using uniform priors for $M_1$ and $M_2$.)}
    \label{tab:inspiral_var_params}
\end{table*}

\begin{table*}
    \centering
    {\renewcommand{\arraystretch}{1.3}
    \begin{tabular}{||c|c||c|c||}
        \hhline{|t:==:tt:==:t|}
         Parameter & Description & Prior distribution & Prior range \\
        \hhline{|:==::==:|}
         $M_f$ & Source frame final mass & Uniform & $[10, 200] M_\odot$ \\
         \hhline{||--||--||}
         $\chi_f$ & Final spin & Uniform & $[-0.99, 0.99]$ \\
         \hhline{||--||--||}
         $A_{220}$ & Initial $(2,2,0)$ mode amplitude & $\log_{10}$ & $[10^{-25}, 8\times 10^{-17}]$ \\
         \hhline{||--||--||}
         $\phi_{220}$ & $(2,2,0)$ mode phase & Uniform & $[0, 2\pi]$ \\ 
         \hhline{||--||--||}
         $A_{221}/A_{220}$ & Initial $(2,2,1)$ mode amplitude (as ratio of $A_{220}$) & Uniform & $[0, 5]$ \\ 
         \hhline{||--||--||}
         $\phi_{221}$ & $(2,2,1)$ mode phase & Uniform & $[0, 2\pi]$ \\
         \hhline{|b:==:bb:==:b|}
    \end{tabular}}
    \caption{Varied parameters in ringdown models and their associated priors. The $t_c$, $\iota$, and sky location priors used in ringdown analyses were identical to those shown in Table \ref{tab:inspiral_var_params}. The amplitude priors indicate the amplitudes of the corresponding quasinormal modes at the start of the ringdown model (i.e.~at the merger). The third column indicates the sampling method for each prior. Here, $\log_{10}$ indicates a uniform distribution over the base 10 logarithm of the given parameter.}
    \label{tab:ringdown_var_params}
\end{table*}

\begin{figure}
    \centering
    \includegraphics[width=0.45\textwidth]{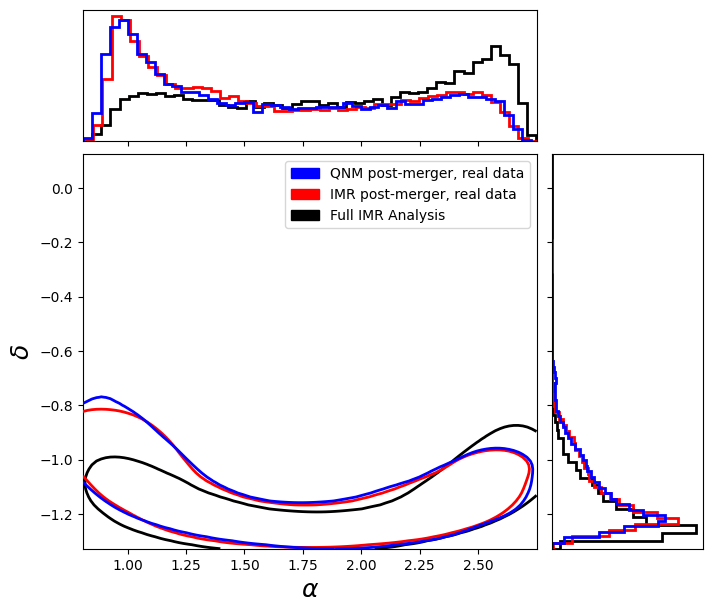}
    \caption{Sky position posteriors for GW150914 analyses. The colored contours represent analyses done using the normalization methods described in Sec.~\ref{sec:det_approx} The blue contour represents the analysis with a QNM postmerger model, and the red contour depicts the analysis with an IMR postmerger model. The black contour represents the posterior from a full IMR analysis conducted in \cite{4ogc}. All contours mark the 90th percentile of each distribution.}
    \label{fig:sky_compare}
\end{figure}

As a preliminary test, the sky location posteriors of the analyses with variable sky/$t_c$ parameters on real data were plotted. As seen in Fig.~\ref{fig:sky_compare}, the analyses were able to recover most of the posterior for the full IMR analysis conducted in \cite{4ogc}.

\section{Area theorem} \label{sec:area_thrm}

The simplest area theorem test is to compare the progenitor horizon areas $A_1$ and $A_2$ to the remnant horizon area $A_f$ to check that
\begin{equation}
    \label{eqn:area_thorem}
    A_1 + A_2 \equiv A_i \leq A_f.
\end{equation}
In natural units ($G=c=1$) the area of each black hole is given by~\cite{area-theorem}:
\begin{equation}
    \label{eqn:bharea}
    A = 8\pi M^2 (1 + \sqrt{1-\chi^2}),
\end{equation}
where $M$ the black hole's mass and $\chi = J/M^2$ is its dimensionless spin.

A more robust test can be performed by comparing the measured change in area to the expected change in area~\cite{kastha-area-thrm}
\begin{equation}
    \label{eqn:delta_area}
    H = \frac{A_{f, \rm measured} - A_i}{A_{f, \rm expected} - A_i}.
\end{equation}
Here, $A_{f, \rm measured}$ \proof{is the area} of the final black hole inferred from the postmerger analysis and $A_i$ is the sum of the initial areas inferred from the premerger analysis. The expected area of the final black hole $A_{f, \rm expected}$ is derived from the initial masses and spins measured from the premerger analysis. These are converted into an estimated final mass and spin using fits from numerical relativity, which is then converted into an area via Eq.~\eqref{eqn:bharea}.

Since $A_{f, expected}$ is evaluated with numerical relativity using $A_1$ and $A_2$, the denominator of Eq.~\eqref{eqn:delta_area} is positive definite. Therefore, if $H < 0$, then $A_{f, measured} < A_i$ and the area theorem is violated. The confidence interval in favor of the area theorem is the fraction of the posterior for which $H > 0$. More robustly, $H=1$ indicates that signal is consistent with GR specifically, not just the more generic class of theories that satisfy the area increase law.

\begin{figure}
    \centering
    \includegraphics[width=0.45\textwidth]{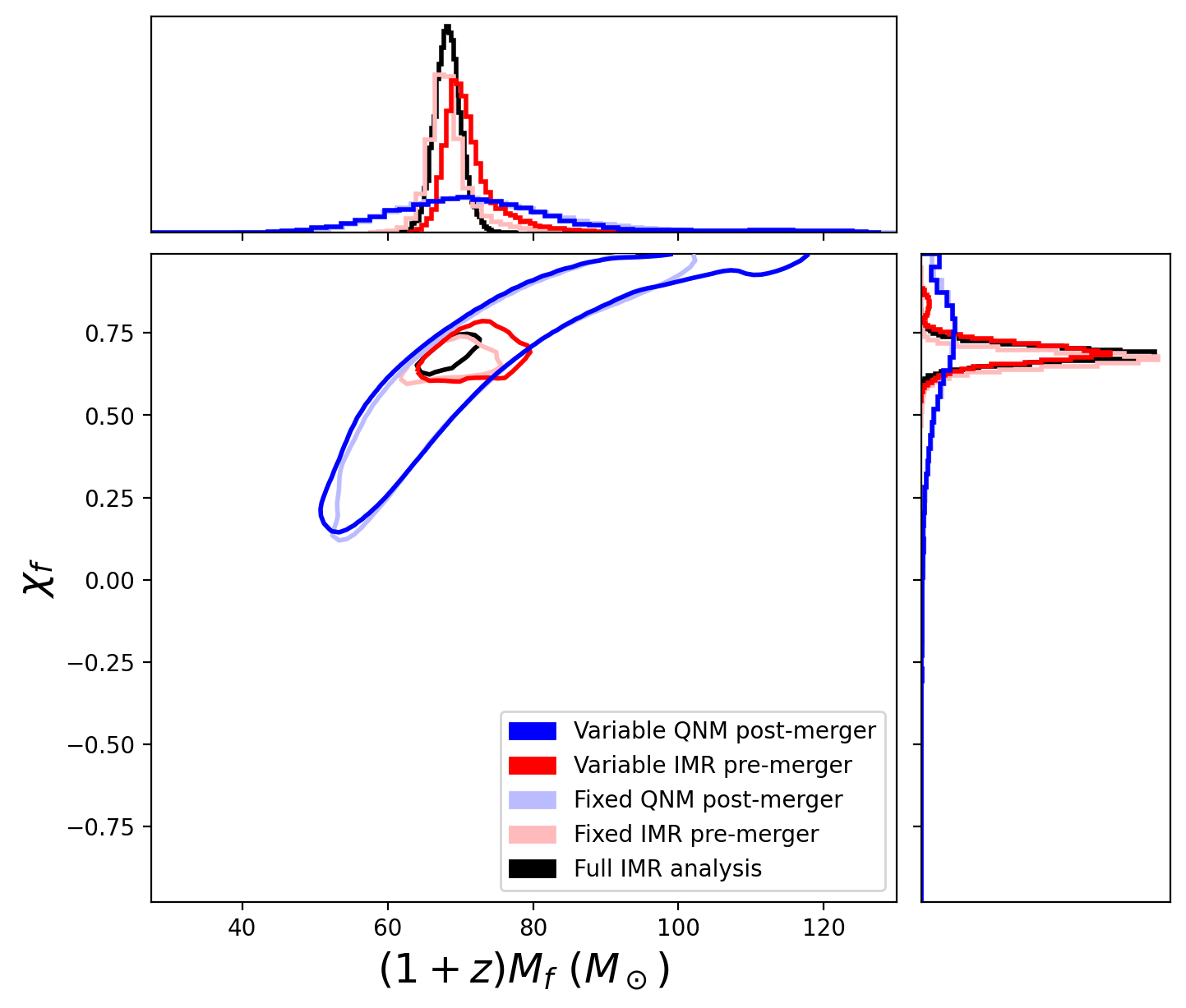}
    \caption{Final mass and final spin posteriors of GW150914 analyses using a QNM postmerger approximant and real waveform data. All masses are in the detector frame. Contours represent the 90th percentile of each distribution. The darker colored lines correspond to the model with variable sky location and $t_c$, and the lighter lines represent the model where these parameters were fixed. Blue lines correspond to postmerger results, while red lines correspond to premerger results, obtained using numerical relativity with the IMR parameters. The innermost black contour represents the 4-OGC posterior from \cite{4ogc}.}
    \label{fig:chi_m_real_qnm}
\end{figure}

Figure~\ref{fig:chi_m_real_qnm} shows the posteriors of the final mass $M_f$ and final spin $\chi_f$ of the QNM postmerger models on real data. Specifically, the $M_f$ and $\chi_f$ posteriors from the analysis with variable sky/$t_c$ parameters are compared with the corresponding posteriors from fixed gate analysis. All posteriors contain within them the distribution from a full IMR analysis from 4-OGC \cite{4ogc}. Furthermore, both posteriors from the variable sky/$t_c$ run overlap almost completely with their corresponding fixed parameter posteriors.

\begin{figure}
    \centering
    \includegraphics[width=0.45\textwidth]{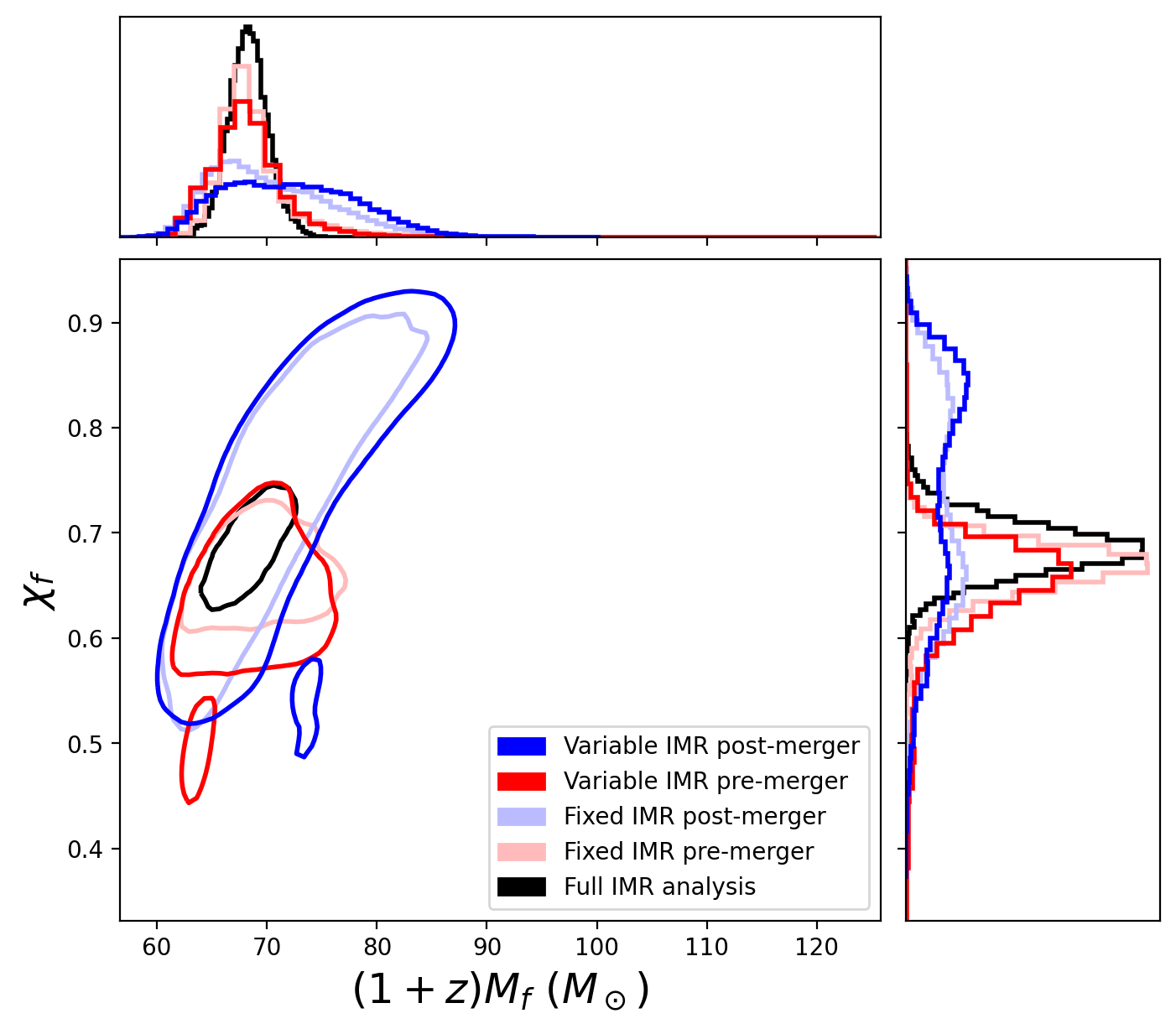}
    \caption{Final mass and final spin posteriors of GW150914 analyses using an IMR postmerger approximant and real waveform data. All masses are in the detector frame. Contours represent the 90th percentile of each distribution. The darker colored lines correspond to the model with variable sky location and $t_c$, and the lighter lines represent the model where these parameters were fixed. Blue lines correspond to postmerger results, while red lines correspond to premerger results, obtained using numerical relativity with the IMR parameters. The innermost black contour represents the 4-OGC posterior from \cite{4ogc}.}
    \label{fig:chi_m_real_imr}
\end{figure}

A similar plot is shown in Fig.~\ref{fig:chi_m_real_imr}, except with the analyses that used an IMR postmerger model on real data. Again, the premerger and postmerger posteriors from the analysis with variable sky/$t_c$ parameters are compared to their fixed counterparts as well as the full IMR posterior from \cite{4ogc}. Notably, while still in very close agreement, the variable sky/$t_c$ posteriors have a greater discrepancy from their fixed counterparts than seen in Fig.~\ref{fig:chi_m_real_qnm}. Specifically, the postmerger posterior tends towards slightly higher masses and spins, while the premerger posterior includes lower spin values. 

Directly comparing the premerger posteriors in Figs.\ref{fig:chi_m_real_qnm} and \ref{fig:chi_m_real_imr} shows a slight discrepancy in final mass and spin estimates. Namely, the analysis with a QNM postmerger model tends towards higher final mass and spin estimates than the IMR postmerger model. This indicates minor coupling between the pre and postmerger models caused by their shared sky and time parameters. However, as made evident by these two plots, this is not a major effect, and both premerger models maintain strong consistency with the full IMR posterior.

\begin{figure}
    \centering
    \includegraphics[width=0.45\textwidth]{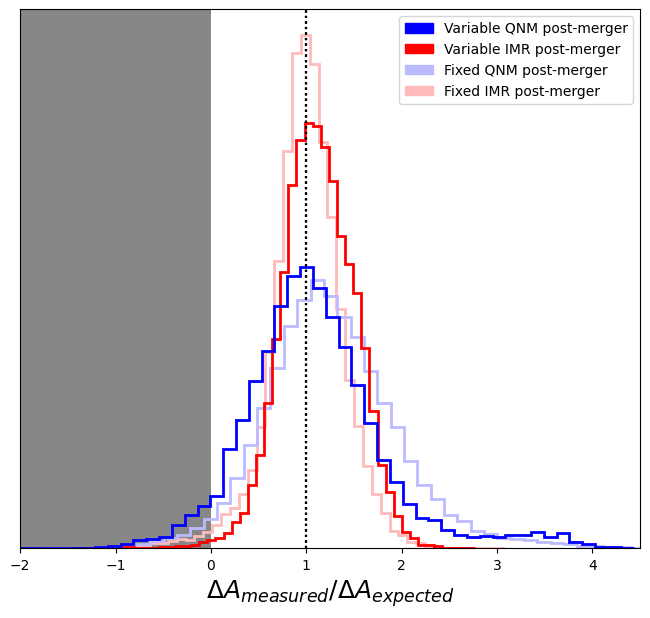}
    \caption{Posteriors of $H = (A_{f, measured} - A_i)/(A_{f, expected} - A_i)$ for the GW150914 analyses with real waveform data. Blue histograms correspond to QNM postmerger models, while red lines correspond to IMR postmerger models. Darker histograms correspond to models with variable sky/$t_c$ parameters, while lighter histograms correspond to models where these parameters are fixed to the maximum likelihood values from \cite{4ogc}. The expected value $H = 1$ is shown with a vertical black dotted line. Points in the shaded region correspond to $H < 0$ and therefore disagree with the area theorem.}
    \label{fig:area_real}
\end{figure}

Figure~\ref{fig:area_real} compares the $H$ posteriors between the four analyses of the real GW150914 data. All posteriors agree with the expected value $H = 1$, with each distribution peaking near this value.
While the QNM postmerger posteriors are almost in exact agreement with each other, there is a slight discrepancy between the IMR postmerger posteriors. Specifically, the fixed sky/$t_c$ posterior has a much sharper peak at a lower value than the variable gate counterpart. This may be explained by Fig.~\ref{fig:chi_m_real_imr}, where the fixed IMR postmerger posterior was observed to contain lower masses and spins than the variable gate model. This could lead to the slight bias to lower $H$ values seen here.

Table \ref{tab:area_thrm_results} summarizes the $H$ credible intervals and area theorem confidence intervals for all eight analyses. All credible intervals are in agreement with the expected value $H = 1$. Additionally, all analyses of the real GW150914 signal have $H>0$ at greater than 95\% confidence, indicating very high agreement with the area theorem. Results from the analysis of the simulated signal are largely consistent with these results (see Appendix~\ref{sec:addtl_results} for equivalent figures).

The area theorem confidence interval of any given analysis appears to be uncorrelated to whether or not the sky location and coalescence time were allowed to vary. However, IMR postmerger models tended to have slightly higher agreement with the area theorem than corresponding QNM postmerger models. This pattern may be explained by trends in the pre and postmerger mass and spin results. The IMR postmerger models in Fig.~\ref{fig:chi_m_real_imr} had postmerger measurements that were either in agreement with or greater than the corresponding premerger estimates. This corresponds to a very high concentration of $H$ measurements approximating or exceeding 1, with very few points skewed to negative $H$ values. Conversely, the postmerger posteriors shown in Fig.~\ref{fig:chi_m_real_qnm} have significant amounts of points with higher and lower mass and spin values than their premerger counterparts. This leads to a wider distribution of $H$ values to both higher and lower values, allowing for a higher concentration of negative measurements. 

\begin{table*}
    \centering
    {\renewcommand{\arraystretch}{1.3}
    \begin{tabular}{||c||c|c||}
        \hhline{|t:=:t:==:t|}
         GW150914 analysis & Credible interval & $P(H > 0)$ \\
        \hhline{|:=::==:|}
         Variable sky/$t_c$, QNM postmerger, real waveform & $1.0^{+1.4}_{-1.0}$ & 95.4\% \\
         \hhline{||-||--||}
         Variable sky/$t_c$, IMR postmerger, real waveform & $1.1^{+0.6}_{-0.6}$ & 99.5\% \\
         \hhline{||-||--||}
         Fixed sky/$t_c$, QNM postmerger, real waveform & $1.2^{+1.2}_{-1.0}$ & 97.4\% \\
         \hhline{||-||--||}
         Fixed sky/$t_c$, IMR postmerger, real waveform & $1.0^{+0.6}_{-0.6}$ & 98.4\% \\
         \hhline{||-||--||}
         Variable sky/$t_c$, QNM postmerger, zero-noise injection & $0.9^{+1.5}_{-1.3}$ & 89.1\% \\
         \hhline{||-||--||}
         Variable sky/$t_c$, IMR postmerger, zero-noise injection & $0.9^{+0.6}_{-0.6}$ & 98.4\% \\
         \hhline{||-||--||}
         Fixed sky/$t_c$, QNM postmerger, zero-noise injection & $1.1^{+0.9}_{-0.9}$ & 97.3\% \\
         \hhline{||-||--||}
         Fixed sky/$t_c$, IMR postmerger, zero-noise injection & $1.0^{+0.5}_{-0.5}$ & 98.8\% \\
         \hhline{|b:=:b:==:b|}
    \end{tabular}}
    \caption{Summary of posteriors for $H = (A_{f, measured} - A_i)/(A_{f, expected} - A_i)$ from various GW150914 analyses. The first column lists the properties of each model, namely whether or not sky location and $t_c$ were allowed to vary, the postmerger model used, and the waveform data used. The approximant \texttt{IMRPhenomXPHM} \cite{imrphenomxphm} was used to model the postmerger signals of analyses labeled ``IMR postmerger" as were all premerger signals. Analyses with injected waveforms used a zero-noise injection of the GW150914 waveform as input data. The second column lists the median and the 90\% credible interval for each $H$ posterior. The third column lists the percentage of posterior points for which $H > 0$, thereby agreeing with the area theorem.}
    \label{tab:area_thrm_results}
\end{table*}


\section{Conclusions} \label{sec:conclusions}

This paper presented a novel method of marginalizing over sky location and coalescence time when performing a pre or postmerger analysis, which allows for a full accounting of uncertainties in parameter estimates. Tests of the area theorem were conducted using data from GW150914 using this method. It was found that the data for this event agrees very well with the area theorem regardless of whether sky position and $t_c$ were allowed to vary. The only noticeable changes in agreement in the area theorem were caused by the postmerger approximant used, though in general this was a very minor change.

While we focus on tests of the area theorem here, our method for marginalizing over sky location and $t_c$ applies to any analysis in which only a portion of the signal is modelled. In particular, black hole spectroscopy involves analyzing the postmerger signal using a QNM template in order to perform a test of the no-hair theorem~\cite{Vishveshwara:1970zz,Chandrasekhar:1975zza,Dreyer:2003bv}. As with previous tests of the area theorem, previous black hole spectroscopy studies have fixed the sky location and $t_c$ for the reasons discussed in Sec.~\ref{sec:det_approx}~\cite{isi-221-initial,cotesta-221,LIGOScientific:2021sio}. Our finding that a signal model is needed for the entire observable signal applies equally well to black hole spectroscopy, even though the premerger signal is purposely excluded in such analyses. The hierarchical method we develop in Sec.~\ref{sec:challenges} is therefore equally relevant for marginalizing over sky location and $t_c$ when doing black hole spectroscopy. In Ref.~\cite{Correia:2023bfn} we use our QNM analysis of GW150914 to investigate the evidence for the presence of the $(2,2,1)$ mode, which has been hotly contested in the literature~\cite{isi-221-initial,cotesta-221,isi-221-rebuttal,Wang:2023xsy}.

In all our analyses here we end the premerger analysis when the postmerger analysis beings. This is necessitated by our finding that some model must exist for the entire observable signal when the gate time is allowed to vary. Since there was no gap between our pre and postmerger template, we obtained agreement with the area theorem in excess of $95\%$. A more rigorous test of the area theorem is to excise the merger from the data, since any biases introduced by including the merger as part of the premerger model would be omitted. Refs.~\cite{isi-area-thrm,kastha-area-thrm} did this additional test with fixed sky location and $t_c$. However, this is not possible when marginalizing over sky location and $t_c$ for the reasons highlighted in Sec.~\ref{sec:challenges}. Introducing a gap between the pre and postmerger will once again favor points that excise as much of the signal as possible, even when both the pre and postmerger were modeled simultaneously.

Introducing a gap between the pre and postmerger models could be achieved by using three subdomains rather than two: one each for the inspiral, merger, and ringdown. The gate for one domain would start/end at the end/start of the next. For the merger domain, an arbitrary signal model using wavelets could be used, similar to what Finch and Moore used in Ref.~\cite{finch-moore}. This would ensure that pre and postmerger parameters are fully unbiased by the merger while ensuring that the merger itself is not arbitrarily gated out. The initial black hole areas could then be estimated using an inspiral model and the final area using a QNM model. This would also be useful in black hole spectroscopy studies involving fundamental angular QNMs. These modes are not expected to become relevant until $\sim10 M$ after merger, necessitating a gap between the merger and the start of the QNM model. We plan to investigate this in a future study.

This research was conducted using PyCBC~\cite{pycbc-software}. Our data is available at~\cite{data-release}.

\section*{Acknowledgements} \label{sec:ack}

A.C. was supported by funds from the Massachusetts Space Grant Consortium. C.C. acknowledges support from NSF Grant No.~PHY-2309356.
All computations were performed on Unity, a collaborative, multi-institutional high-performance computing cluster managed by UMass Amherst Research Computing and Data.

This research has made use of data or software obtained from the Gravitational Wave Open Science Center (\footnote{\href{gwosc.org}{gwosc.org}}), a service of the LIGO Scientific Collaboration, the Virgo Collaboration, and KAGRA. This material is based upon work supported by NSF's LIGO Laboratory which is a major facility fully funded by the National Science Foundation, as well as the Science and Technology Facilities Council (STFC) of the United Kingdom, the Max-Planck Society (MPS), and the State of Niedersachsen/Germany for support of the construction of Advanced LIGO and construction and operation of the GEO600 detector. Additional support for Advanced LIGO was provided by the Australian Research Council. Virgo is funded, through the European Gravitational Observatory (EGO), by the French Centre National de Recherche Scientifique (CNRS), the Italian Istituto Nazionale di Fisica Nucleare (INFN) and the Dutch Nikhef, with contributions by institutions from Belgium, Germany, Greece, Hungary, Ireland, Japan, Monaco, Poland, Portugal, Spain. KAGRA is supported by Ministry of Education, Culture, Sports, Science and Technology (MEXT), Japan Society for the Promotion of Science (JSPS) in Japan; National Research Foundation (NRF) and Ministry of Science and ICT (MSIT) in Korea; Academia Sinica (AS) and National Science and Technology Council (NSTC) in Taiwan.

\appendix

\section{LIKELIHOOD FUNCTION DERIVATION} \label{sec:likelihood_der}

To calculate the posterior from Bayes' theorem [Eq.~\eqref{eqn:bayes_thrm}], one requires a model for both the signal $h$ and the noise $n$. While $h$ can be determined from the linearized EFE wave solution, $n$ requires more statistical considerations.

To start, consider a network of $K$ gravitational wave detectors that each sample at a rate $\Delta t$ over a total time length $T$. Defining the total number of samples $N = T/\Delta t$, the full data series $\boldsymbol{s}_{net}$ can be expressed as a series of $K$ $N$-dimensional vectors $\boldsymbol{s}^K$ such that $\boldsymbol{s}^K = \{s^K_0, s^K_1,...,s^K_N\}$ and $\boldsymbol{s}_{net} = \{ \boldsymbol{s}^1, \boldsymbol{s}^2,...,\boldsymbol{s}^K \}$. To simplify calculations, assume that the signal is zero such that $\boldsymbol{s}_{net} = \boldsymbol{n}$. Additionally, assume that noise model is a stochastic Gaussian distribution and is uncorrelated between detectors. Under these assumptions, the noise likelihood function is
\begin{equation}
    p(\boldsymbol{s}_{net} | n) = \frac{\exp{[-\frac{1}{2} \sum_{d=1}^K \boldsymbol{s}^T_d \boldsymbol{\Sigma}^{-1}_d \boldsymbol{s}_d]}}{[(2\pi)^{NK} \prod_{d=1}^K \det \boldsymbol{\Sigma}_d]^{1/2}},
    \label{eqn:det_likelihood}
\end{equation}
where $\boldsymbol{\Sigma}_d$ is the covariance matrix of the noise model for detector $d$, defined using the ensemble average:
\begin{equation}
    \boldsymbol{\Sigma}_d [j, k] = \langle \boldsymbol{s}_d[j] \boldsymbol{s}_d[k] \rangle.
    \label{eqn:cov_matrix}
\end{equation}
(By dropping the $d$ subscripts, one obtains an equivalent expression for the full covariance matrix and data. We do so in the following steps for brevity.) This is the exact expression of the likelihood function for noise. However, this function is infeasible to calculate analytically due to the inverse covariance matrix in the numerator.

To do this, one may expand the covariance matrix definition in Eq.~\eqref{eqn:cov_matrix} as follows, defining $\Delta_{kj} = k-j$:
\begin{align}
    \boldsymbol{\Sigma} [j, k] &= \langle \boldsymbol{s}[j] \boldsymbol{s}[k] \rangle \nonumber \\
    &= \langle \boldsymbol{s}[j] \boldsymbol{s}[\Delta_{kj}+j] \rangle \nonumber \\
    &= \lim_{n\to\infty} \frac{1}{n} \sum^{n-1}_{l=0} s^l[j]s^l[\Delta_{kj}+j],
    \label{eqn:expand_cov}
\end{align}
where in the last step the ensemble average is written out fully. 

In general, this expression is dependent on time $t_j = j\Delta t$ and displacement $\tau_{kj} = \Delta_{kj}\Delta t$. However, one can make the assumption that the noise is wide sense stationary, where the mean and variance are both constant in time. Under this assumption, any constant can be added to the indices in Eq.~\eqref{eqn:cov_matrix} to obtain the same result. This makes $\boldsymbol{\Sigma}$ symmetric [since the factors in Eq.~\eqref{eqn:expand_cov} commute] and Toeplitz (since the elements along the diagonals are equal) \cite{toeplitz-circulant}. Additionally, since $\boldsymbol{\Sigma} [0, \Delta_{kj}] = \boldsymbol{\Sigma} [-\Delta_{kj}, 0] = \boldsymbol{\Sigma} [0, -\Delta_{kj}]$, the elements of $\boldsymbol{\Sigma}$ are even functions of $\Delta_{kj}$.

Additionally, one can assume that the data is ergodic, meaning that new realizations of $\boldsymbol{s}$ are obtained via time. Under this assumption and the properties of the elements of $\boldsymbol{\Sigma}$, the ensemble averages in Eq.~\eqref{eqn:expand_cov} can be replaced with time averages,
\begin{align}
    \boldsymbol{\Sigma} [j, k] &= \lim_{n\to\infty} \frac{1}{n} \sum^{n-1}_{l=0} s^l[0]s^l[\Delta_{kj}] \nonumber \\
    &= \lim_{n\to\infty} \frac{1}{n} \sum^{n-1}_{l=0} s^l[l]s^l[\Delta_{kj}+l] \nonumber \\
    &= \lim_{n\to\infty} \frac{1}{2n} \sum^{n-1}_{l=-n} s^l[l]s^l[\Delta_{kj}+l] \nonumber \\
    &= \frac{1}{2} R_{ss}((k-j)\Delta t).
    \label{eqn:cov_rss}
\end{align}
The last step defines the autocorrelation function $R_{ss}(\tau)$, which describes the correlation between points in the time series $\boldsymbol{s}$. If $R_{ss}(\tau)$ goes to zero in some finite time $\tau_{max}$, then all diagonals with $|\Delta_{kj}| > floor(\tau_{max}/\Delta t) = \Delta_{max}$ will equal zero. This is similar to the form of a circulant matrix $\boldsymbol{C}$, a special case of a Toeplitz matrix where each row is a right-cycle permutation of the same vector (in this case, $\boldsymbol{s}$). The eigenvectors of circulant matrices are well-known \cite{toeplitz-circulant},
\begin{equation}
    \boldsymbol{u}_p[k] = \frac{1}{\sqrt{N}}\exp{(-2\pi ikp/N)}.
    \label{eqn:circ_eigenvals}
\end{equation}
This generally is not true for Toeplitz matrices, but \proof{one may take advantage of Eq.~\eqref{eqn:circ_eigenvals}} by recognizing that the matrix described by Eq.~\eqref{eqn:cov_rss} asymptotes to a circulant matrix for large N,
\begin{equation}
    \lim_{N\to\infty} |\boldsymbol{\Sigma} - \boldsymbol{C}| = \boldsymbol{0}.
    \label{eqn:cov_circ_lim}
\end{equation}

Therefore, the eigenvalues $\lambda_p$ of $\boldsymbol{\Sigma}$ can be evaluated using the usual eigenvalue equation as long as $\Delta_{max} << N/2$,
\begin{equation}
    \boldsymbol{\Sigma} \boldsymbol{u}_p \approx \lambda_p \boldsymbol{u}_p.
    \label{eqn:toep_eigenvals}
\end{equation}

Therefore, using the fact that $\boldsymbol{\Sigma}$ is symmetric and $R_{ss}(l)$ is even,
\begin{align}
    \lambda_p &= \frac{1}{2} Re\bigg\{ \sum^{N/2}_{l=-N/2} R_{ss}(l)\exp (-2\pi ipl/N) \bigg\} \nonumber \\
    &= \frac{1}{2} Re\bigg\{ \sum^{N-1}_{l=0} R_{ss}(l)\exp (-2\pi ipl/N) \bigg\} \nonumber \\
    &= \frac{1}{2} Re\{ \tilde{R}_{ss}(p) / \Delta t \},
    \label{eqn:cov_rss_eigenvals}
\end{align}
where $\tilde{R}_{ss}(p)$ is the discrete Fourier transform of the autocorrelation function,
\begin{equation}
    \tilde{R}_{ss}(p) = \Delta t \sum^{N-1}_{k=0} {R}_{ss}(k) \exp (-2\pi ipk/N).
    \label{eqn:tilde_Rss}
\end{equation}

To simplify Eq.~\eqref{eqn:cov_rss_eigenvals}, one may impose the Wiener-Khinchin Theorem \cite{time-series}, which defines the power spectral density $S_n$ as the Fourier transform of $R_{ss}$ for a wide-sense stationary stochastic process,
\begin{equation}
    \lambda_p = \frac{S_n[p]}{2\Delta t}
    \label{eqn:wk_thrm}
\end{equation}

One can then construct the inverse of $\boldsymbol{\Sigma}$ using the eigenvalues from Eq.~\eqref{eqn:wk_thrm} and the eigenvectors of Eq.~\eqref{eqn:circ_eigenvals},
\begin{align}
    \boldsymbol{\Sigma}^{-1}[j,k] &= \frac{2\Delta t}{N} \sum^{N-1}_{p=0} \frac{\exp [-2\pi ip(j-k)/N]}{S_n(p)} \nonumber \\
    &= {2\Delta f (\Delta t)^2} \sum^{N/2-1}_{p=0} \frac{e^{-2\pi ip(j-k)/N} + e^{2\pi ip(j-k)/N}}{S_n(p)},
    \label{eqn:cov_inverse}
\end{align}
using the fact that $S_n$ is symmetric about $N/2$ and defining the sample frequency $\Delta f = 1/T = 1/N\Delta t$. Since $\boldsymbol{s}$ is real and
\begin{equation}
    (\Delta t)^2\sum^{N-1}_{j,k=0} \boldsymbol{s}[j] \boldsymbol{s}[k](e^{-2\pi ip(j-k)/N} + e^{2\pi ip(j-k)/N}) = 2|\tilde{\boldsymbol{s}}|^2[p],
    \label{eqn:real_data_identity}
\end{equation}
one can write
\begin{equation}
    \boldsymbol{s}^T \boldsymbol{\Sigma}^{-1} \boldsymbol{s} = 4\Delta f \sum^{N/2-1}_{p=0} \frac{2|\tilde{\boldsymbol{s}}|^2[p]}{S_n(p)}.
    \label{eqn:inv_cov_prelim}
\end{equation}

Finally, if the inner product between two arbitrary vectors $\boldsymbol{a}$ and $\boldsymbol{b}$ is defined as
\begin{equation}
    \langle \boldsymbol{a}, \boldsymbol{b} \rangle = 4Re\bigg\{ \Delta f \sum^{N/2-1}_{p=0} \frac{\tilde{\boldsymbol{a}}^{\*}[p]\tilde{\boldsymbol{b}}[p]}{S_n(p)} \bigg\},
    \label{eqn:inner_product}
\end{equation}
Eq.~\eqref{eqn:inv_cov_prelim} can be written as an inner product,
\begin{equation}
    \boldsymbol{s}^T \boldsymbol{\Sigma}^{-1} \boldsymbol{s} = \langle \boldsymbol{s},  \boldsymbol{s} \rangle,
    \label{eqn:inv_cov_inner_prod}
\end{equation}
and the likelihood function can be expressed as
\begin{equation}
    p(\boldsymbol{s}_{net} | n) = \frac{\exp{[-\frac{1}{2} \sum_{d=1}^K \langle \boldsymbol{s}_d,  \boldsymbol{s}_d \rangle]}}{[(2\pi)^{NK} \prod_{d=1}^K \det \boldsymbol{\Sigma}_d]^{1/2}}
    \label{eqn:likelihood_inner_prod}
\end{equation}

\section{GATING AND IN-PAINTING}  \label{sec:gate_paint}

Generally, BBH models do not require the entire waveform to be analyzed at once. For example, the analyses throughout this paper independently examine the pre and postmerger portions of the GW150914 waveform. To maintain independence between the models, any points not corresponding to the respective model (i.e.~after $t_c$ for the premerger model, or before $t_c$ for the postmerger model) were excised, or ``gated," from the data. Here, a gate of length $M$ applied starting at a sample $a$ will excise all samples within the range $[a, a+M]$, corresponding to a gate of time length $t = M\Delta t$. The gated time series will therefore take the form $\boldsymbol{s}_{d,tr} = \{ s_0, s_1, ... s_a, s_{a+M}, ... s_{N-1}, s_N \}$. 

By doing this, the simplifications made to derive the likelihood function are no longer valid, since $\boldsymbol{\Sigma}_{d,tr}$ is no longer Toeplitz (and, subsequently, no longer approximately circulant for large $N$). There are numerical methods to calculate the matrix inverse directly, but in general they can be unstable and time intensive. Therefore, a method known as ``gating and in-painting" is employed for gated waveform analyses \cite{gating-inpainting}.

First, the method assumes that the noise time series $\boldsymbol{n}$ is the sum of $\boldsymbol{n}_g$, the noise series with the gated times zeroed out, and $\boldsymbol{x}$, a time series containing only the gated samples in $\boldsymbol{n}$. The goal of in-painting is to solve the following equation in the gated region:
\begin{equation}
    \boldsymbol{\Sigma}^{-1} (\boldsymbol{n}_g + \boldsymbol{x}) = \boldsymbol{0}
    \label{eqn:gate_eqn}
\end{equation}

If the nonzero elements of $\boldsymbol{x}$ are such that $(\boldsymbol{\Sigma}^{-1} \boldsymbol{n})[k] = \boldsymbol{0}$ for all samples $k$ in the gate $[a, a+M]$, then the inner product $\boldsymbol{n}^T \boldsymbol{\Sigma} \boldsymbol{n}$ will be equal for the truncated and raw dataset. Since $\boldsymbol{x}$ is zero outside of the gate, $\boldsymbol{\Sigma} ^{-1} \boldsymbol{x}$ will form an $M \times M$ Toeplitz matrix containing the $[a, a+M]$ rows and columns of $\boldsymbol{\Sigma} ^{-1}$. Therefore, Eq.~\eqref{eqn:gate_eqn} can be rewritten within the gated region as
\begin{equation}
   \boldsymbol{\Sigma}^{-1} \boldsymbol{x} = - \boldsymbol{\Sigma}^{-1} \boldsymbol{n}_g, \label{eqn:inpaint_eqn}
\end{equation}
and adding $\boldsymbol{x}$ to $\boldsymbol{n}_g$ will give the same result as truncating $\boldsymbol{n}$ and $\boldsymbol{\Sigma}$. Unlike trying to solve for the inverse directly, this solution is readily found using a Toeplitz solver \cite{scipy, numpy}. Given gated data $\boldsymbol{s}_g$ containing some gated signal $\boldsymbol{h}_g$, Eq.~\eqref{eqn:inpaint_eqn} can be evaluated using $\boldsymbol{n}_g = \boldsymbol{s}_g - \boldsymbol{h}_g$, and the value $\boldsymbol{x} + \boldsymbol{s}_g - \boldsymbol{h}_g$ can be used to calculate the likelihood.

The analyses conducted in this paper utilize gating and in-painting to apply a gate starting/ending at $t_c$ to the waveform template. In theory, the gates should extend to the edges of the analysis segment (i.e., the premerger gate starts at the segment start time, and the postmerger gate ends at segment end). However, the in-painting algorithm, which is the dominant cost in our analysis, is $\sim\!O(M^2)$ operations for a gate of $M$ samples. Although the observable signal is only $\sim0.2\,$s long~\cite{GW150914-observation}, we use an analysis segment that is $4\,$s in duration in order to resolve line artifacts in the PSD. We also use a sample rate of $2048\,$Hz, in order to fully capture all observable signal power. The in-painting would therefore need to span $M\sim 4096$ samples if the gates were to extend to the beginning/end of the analysis segment. This is computationally expensive: some of our analyses would take $\sim1$ month to complete (utilizing 64 CPU cores).

To reduce the computational cost, we instead use a 1-second gate starting (ending) at $t_c$ and ensuring that the remainder of the premerger (postmerger) template $\boldsymbol{h}$ after (before) the gate is zero. This is equivalent to applying a full gate to the template when a whitening filter is applied, since the early- and late-time template will be identically zero in both cases.

We do not zero the data $\boldsymbol{s}$ outside of the gates, however, since doing so would introduce additional ringing artifacts at the beginning/end of the analysis segment not accounted for by the in-painting. This means that the pre and postmerger models will share data outside of the gates (namely, greater than 1 second before and after $t_c$). However, the data is expected to be noise-dominated in this region, yielding an insignificant contribution to the likelihood. This can be seen in Appendix \ref{sec:addtl_results}, which portrays the remnant posteriors for analyses using zero-noise waveform injections. Since the data in these analyses contain no noise, the shared data has identically zero contribution to the likelihood. These results are in strong agreement with Figs.~\ref{fig:chi_m_real_qnm}, \ref{fig:chi_m_real_imr}, and \ref{fig:area_real}, indicating that any effects due to early- and late-time shared noise are indeed negligible.

\section{SIMULATION RESULTS}  \label{sec:addtl_results}

\begin{figure}
    \centering
    \includegraphics[width=0.45\textwidth]{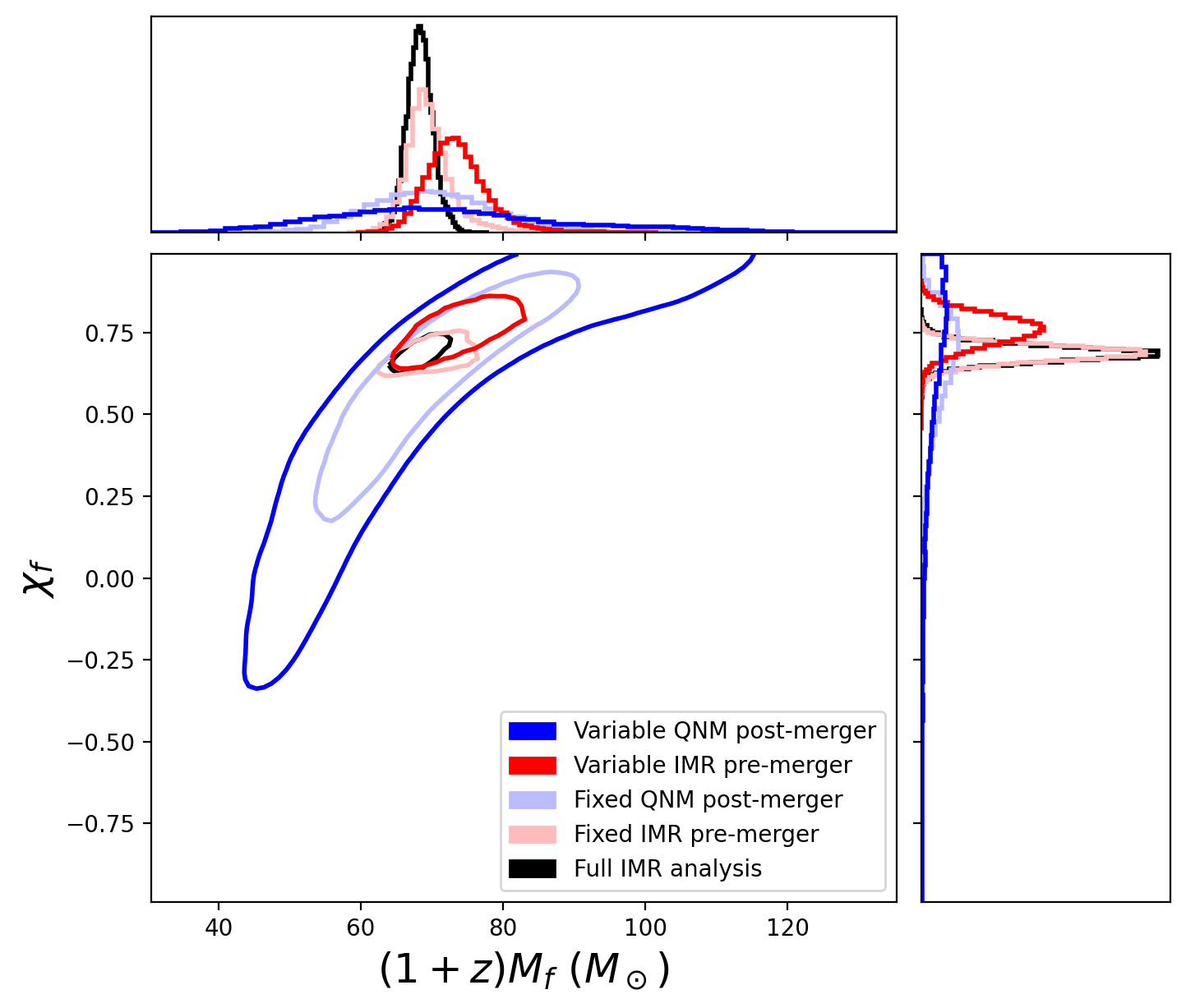}
    \caption{Final mass and final spin posteriors of GW150914 analyses using a QNM postmerger approximant and a zero-noise waveform injection. All masses are in the detector frame. Contours represent the 90th percentile of each distribution. The darker colored lines correspond to the model with variable sky location and $t_c$, and the lighter lines represent the model where these parameters were fixed. Blue lines correspond to postmerger results, while red lines correspond to premerger results, obtained using numerical relativity with the IMR parameters. The innermost black contour represents the 4-OGC posterior from \cite{4ogc}.}
    \label{fig:chi_m_inj_qnm}
\end{figure}

\begin{figure}
    \centering
    \includegraphics[width=0.45\textwidth]{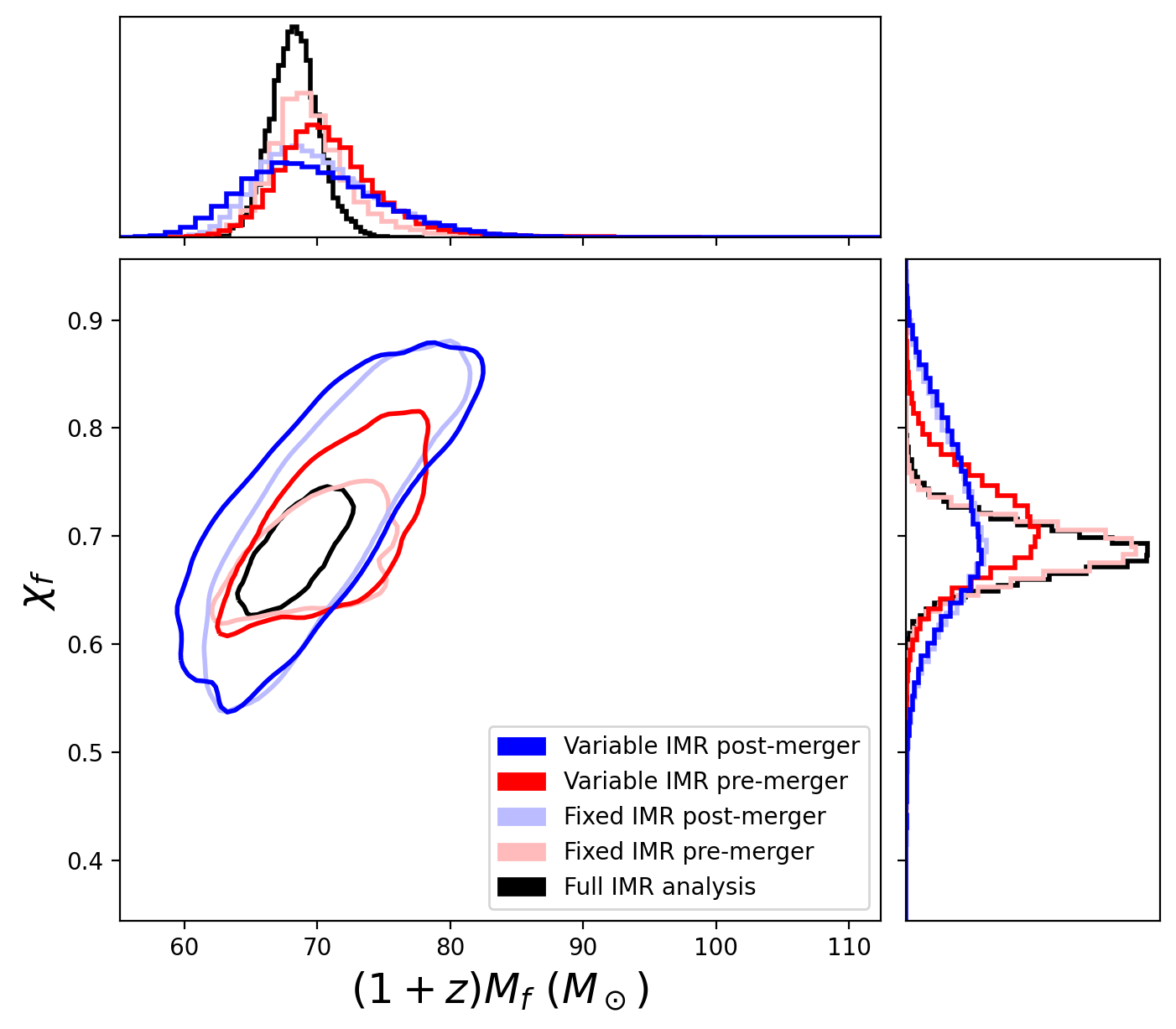}
    \caption{Final mass and final spin posteriors of GW150914 analyses using an IMR postmerger approximant and a zero-noise waveform injection. All masses are in the detector frame. Contours represent the 90th percentile of each distribution. The darker colored lines correspond to the model with variable sky location and $t_c$, and the lighter lines represent the model where these parameters were fixed. Blue lines correspond to postmerger results, while red lines correspond to premerger results, obtained using numerical relativity with the IMR parameters. The innermost black contour represents the 4-OGC posterior from \cite{4ogc}.}
    \label{fig:chi_m_inj_imr}
\end{figure}

\begin{figure}
    \centering
    \includegraphics[width=0.45\textwidth]{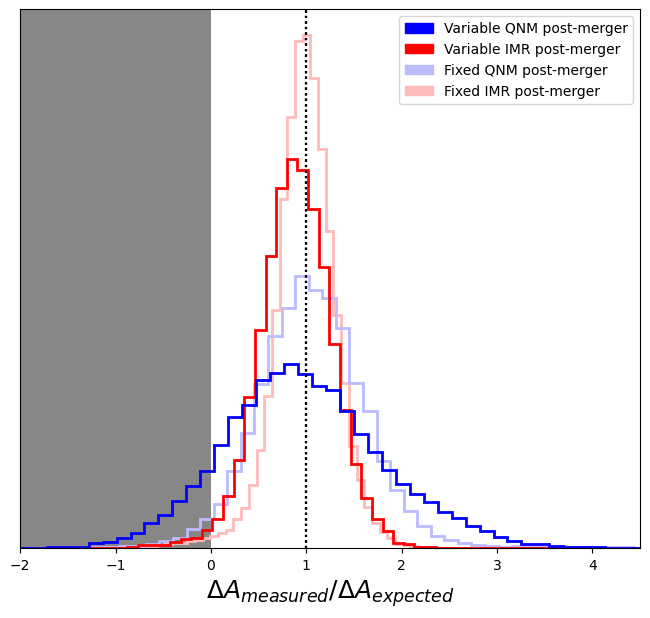}
    \caption{Posteriors of $H = (A_{f, measured} - A_i)/(A_{f, expected} - A_i)$ for the GW150914 analyses with zero-noise injected waveform data. Blue histograms correspond to QNM postmerger models, while red lines correspond to IMR postmerger models. Darker histograms correspond to models with variable sky/$t_c$ parameters, while lighter histograms correspond to models where these parameters are fixed to the maximum likelihood values from \cite{4ogc}. The expected value $H = 1$ is shown with a vertical black dotted line. Points in the shaded region correspond to $H < 0$ and therefore disagree with the area theorem.}
    \label{fig:area_inj}
\end{figure}

This section contains additional plots showing results from the analysis of the GW150914-like simulated signal in zero noise. Figure~\ref{fig:chi_m_inj_qnm} shows the $M_f$ and $\chi_f$ posteriors for the QNM postmerger models, and Fig.~\ref{fig:chi_m_inj_imr} shows the same for the IMR postmerger models. Figure~\ref{fig:area_inj} shows the $H$ posteriors for the zero-noise injection models. 

\begin{figure}
    \centering
    \includegraphics[width=0.45\textwidth]{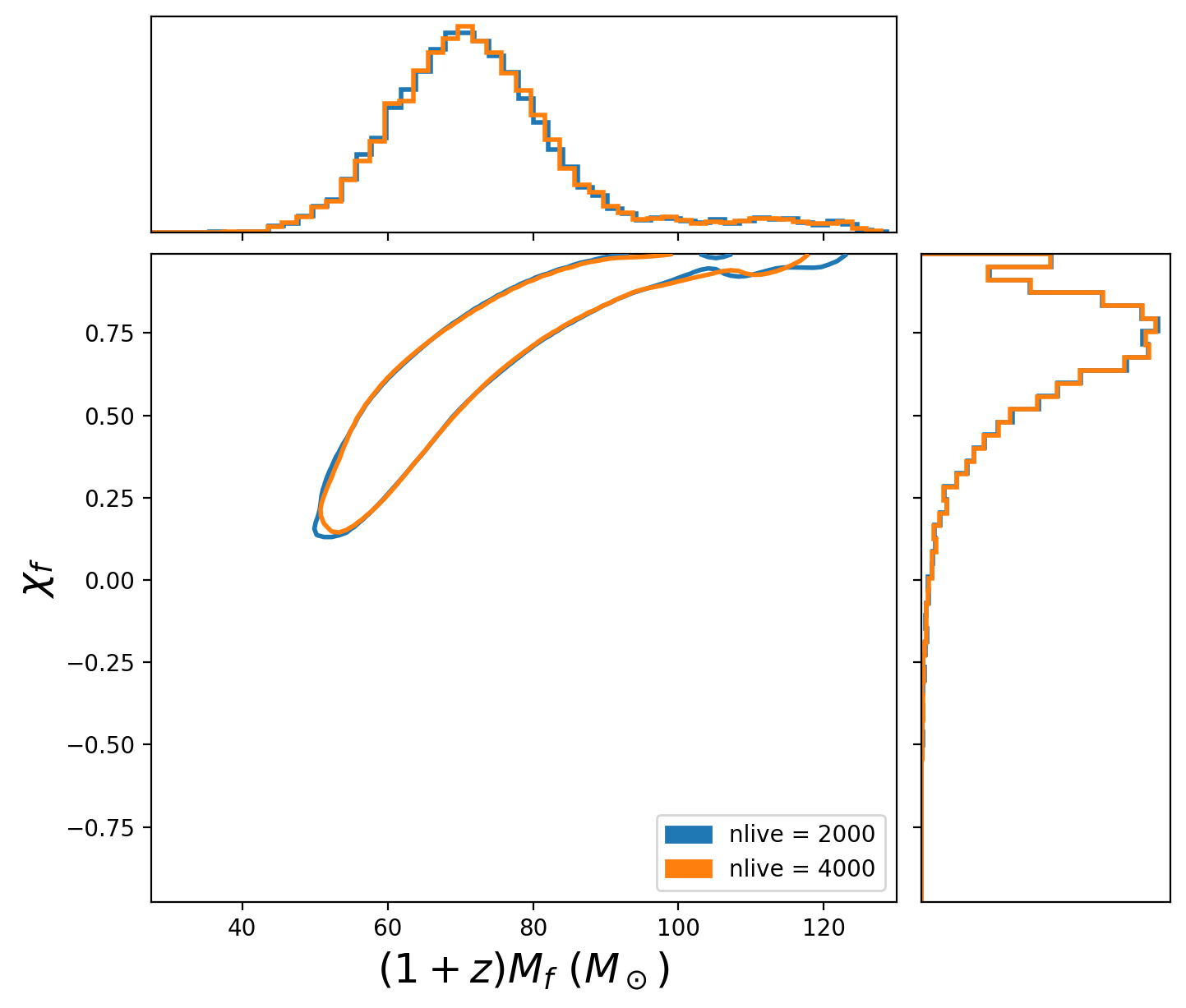}
    \caption{Final mass and spin posteriors for analyses of GW150914 data with a QNM postmerger model and a variable ringdown start time. The blue contour indicates the 90th percentile of an analysis where the \texttt{dynesty} sampler used 2000 live points. The orange contour represents the 90th percentile of an analysis with 4000 live points. Besides the number of sampler points used, both models were identical.}
    \label{fig:qnm_convergence}
\end{figure}

\begin{figure}
    \centering
    \includegraphics[width=0.45\textwidth]{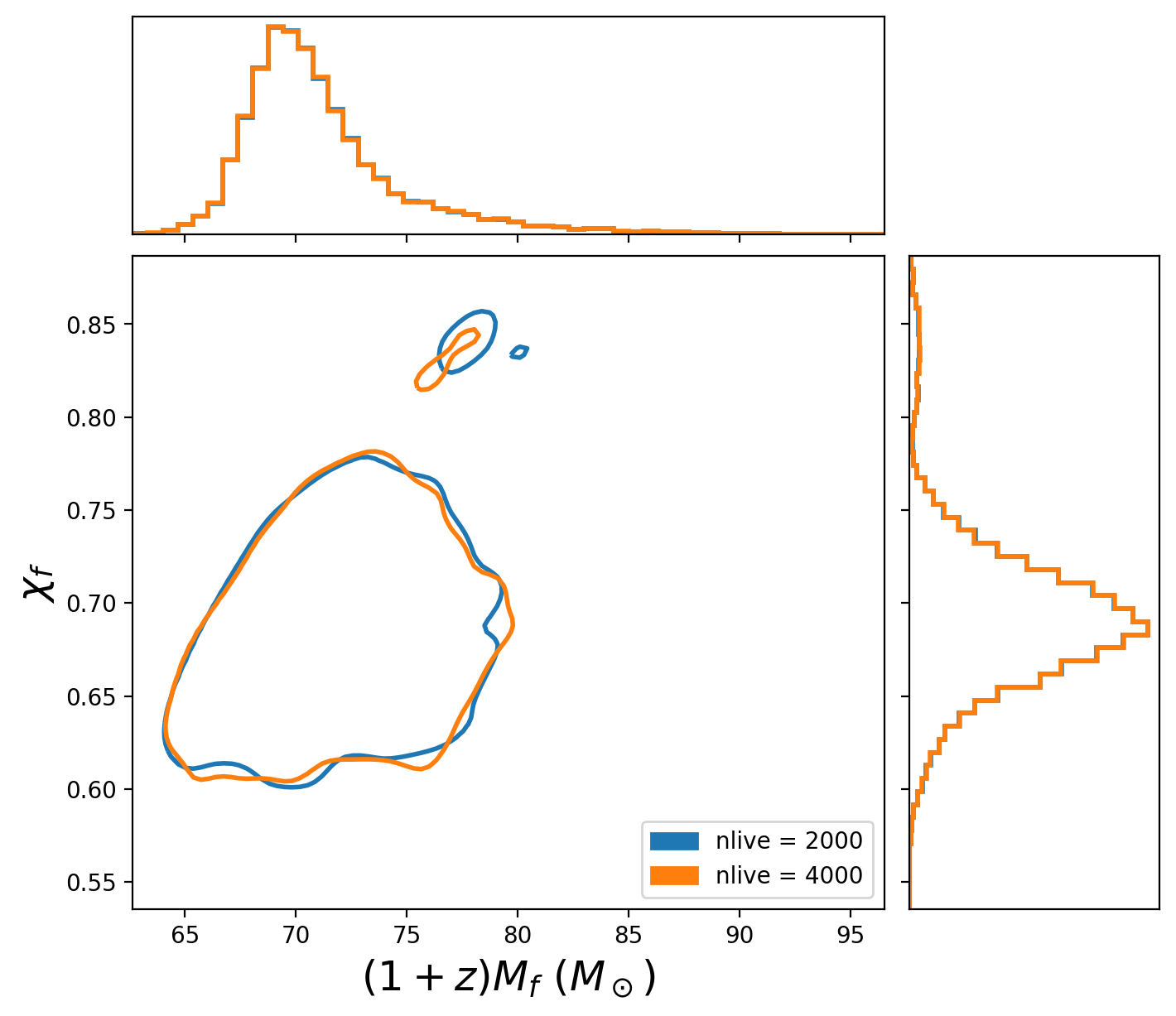}
    \caption{Final mass and spin posteriors for analyses of GW150914 data with a IMR postmerger model and a variable ringdown start time. The blue contour indicates the 90th percentile of an analysis where the \texttt{dynesty} sampler used 2000 live points. The orange countour represents the 90th percentile of an analysis with 4000 live points. Besides the number of sampler points used, both models were identical.}
    \label{fig:imr_convergence}
\end{figure}

\section{SAMPLER CONVERGENCE}  \label{sec:livepoints}

To test the convergence of the sampler we repeated the analyses with 2000 and 4000 live points. Figure~\ref{fig:qnm_convergence} shows the $M_f$ and $\chi_f$ posteriors for QNM postmerger models on real data with 2000 and 4000 live points, while Fig.~\ref{fig:imr_convergence} depicts the same for the IMR postmerger models. Both the QNM and IMR models were able to converge to similar distributions regardless of the number of live points used. All sky/$t_c$ results are reported using 4000 live points.

\bibliography{references-short}

\end{document}